\documentclass[aip,rsi,preprint,groupedaddress]{revtex4-1}

\pdfoutput=1

\usepackage[dvipsnames,svgnames,x11names]{xcolor}
\usepackage[final]{changes}

\usepackage{array, makecell}
\usepackage{dcolumn}   
\usepackage{bm}        
\usepackage{amssymb}   
\usepackage{amsfonts}
\usepackage{verbatim}
\usepackage{siunitx}
\usepackage{amsmath}
\usepackage{amsfonts}
\usepackage{ulem}

\begin{document}
\normalem

\title{High-resolution calorimetric sample platforms for cryogenic thermodynamic studies with multimodal synchrotron x‑ray compatibility}

\author{U. Patel$^{1}$}
\email[Electronic address: ]{upatel@anl.gov}
\author{H. Zheng$^{1}$}
\author{J. L. McChesney$^{1}$}
\author{U. Welp$^{2}$}
\author{Z. Islam$^{1}$}
\author{A. Miceli$^{1}$}

\affiliation{$^{1}$X-ray Science Division, Argonne National Laboratory, Lemont, IL 60439, USA}

\affiliation{$^{2}$Materials Science Division, Argonne National Laboratory, Lemont, IL 60439, USA}

\date{\today}

\begin{abstract}
X-ray calorimetric sample platforms combining specific heat and synchrotron x-ray measurements provide a powerful means to investigate fundamental material properties. Calorimeter cell designs featuring a compact heater and thermometer arranged in a sidecar geometry, with the sample positioned directly above the heater at the center of a silicon nitride membrane, are presented. High-yield, wafer-level batch fabrication of precision calorimetric sensor chips, beamline and laboratory cryostat plugins with sensor mounting and packaging are described. Using our calorimetric sensors, we present specific heat measurements on samples with masses ranging from \SI{4}{\micro\gram} to \SI{145}{\micro\gram}. The sample and reference cells are characterized with relaxation and ac steady-state measurements. The thermal response is captured using lock-in detection at carefully optimized measurement frequencies, with phase-lag correction ensuring precise extraction of heat capacity. The reference cell's background heat capacity was measured to be under 320\,nJ/K at 300\,K, decreasing to just 0.4\,nJ/K at 0.7\,K. The calorimeter performance is illustrated by studying the specific heat of small samples of superconducting Nb and a \SI{4}{\micro\gram} piece of superconducting Al under different magnetic field strengths. The determination of fundamental thermodynamic quantities from low‑temperature electronic and lattice specific heat measurements is discussed. These versatile, high-throughput sample platforms are engineered for small-sample calorimetry across a broad cryogenic temperature range, and they support scalable integration with a wide range of cryostats, including beamline cryostats at the Advanced Photon Source. They accommodate multimodal geometries and enable operation under ultra-high vacuum, millikelvin temperatures, magnetic fields, and x-ray illumination.

\end{abstract}

\maketitle

\section{\added{introduction}}

The emergent phenomena which are the hallmark of quantum materials arises because of competing order parameters from the various degrees of freedom within a material, such as lattice, spin and charge. Understanding how these quantum phases evolve with temperature is paramount in the realization of new devices based on these quantum materials. The heat capacity measurements, particularly at low temperatures where the lattice contributions are small, can be used to identify signatures of quantum phase transitions, collective excitations and other quantum phenomena \cite{Stewart1983}. Coupling x-ray techniques, which can be used to probe the atomic and electronic structure of materials with atomic, electric and spin elemental and orbital specificity, with calorimetry provides a new powerful mean to not only identifying quantum states in matter but understanding the physics governing them. \cite{Willa2017} This multimodal approach helps map complex phase diagrams, uncover hidden states, and follow the interactions between lattice, spin, and electronic properties. Discovering and understanding these new phases not only deepens our knowledge of how matter works but also opens doors to future technologies. 

Over the past several decades, the integration of high-temperature calorimetry particularly fast scanning calorimetry,\cite{schick2016} with x‑ray techniques has advanced significantly. Fast scanning calorimeters apply a continuous, high‑rate temperature ramp, making them particularly effective for probing kinetics and metastable states above room temperature. In the early stages, differential scanning calorimetry was combined with small‑ and wide‑angle x‑ray scattering to reveal ordering phenomena in polymers and biomaterials, primarily through changes in heat capacity.\cite{Russell1985,Clout2016,Lexa2003,Baeten2015,Wurm2009,Bras1995} These studies were limited to milligram‑scale samples due to the large addenda of the calorimeter cells. The subsequent development of thin‑film chip calorimeters enabled faster heating and cooling rates with reduced sample masses, opening the door to nanoscale thermal investigations. Building on these advances, scanning ac nanocalorimetry was later coupled with synchrotron x‑ray diffraction, allowing metallic thin‑film phase transitions to be tracked with simultaneous calorimetric and structural analyses.\cite{Xiao2013,Gregoire2013} Most recently, scanning calorimetry has been integrated with synchrotron total scattering, \cite{Sun2025} diffraction with nano‑focused beams,\cite{Melnikov2016,Rosenthal2014} and x‑ray photon correlation spectroscopy,\cite{Martinelli2024} enabling \textit{in situ} studies of sub‑\text{µg} crystals, thin films, and nanoparticles. Successful integration of complementary techniques, including time-of-flight mass spectrometry\cite{Yi2015} and transmission electron microscopy,\cite{McGieson2025} has also been reported. Despite the importance of cryogenic thermodynamic investigations, low‑temperature calorimetry remains underexplored and could be particularly impactful when applied to quantum materials. This is true both in standalone studies and, even more so, in conjunction with different synchrotron x‑ray methods. The gap arises from the limited availability of high‑resolution cryogenic calorimetric platforms engineered to accommodate small samples while offering multimodal synchrotron x‑ray compatibility. 

In condensed matter physics, specific heat serves as a fundamental thermodynamic characterization tool that applies across many classes of materials. Specific heat measurements are highly valuable for investigating phase transitions-whether structural, magnetic, superconducting, or orbital. In some cases, they provide the only reliable evidence that a transition has occurred as an intrinsic property of the bulk material rather than being caused by a minority phase.\cite{Stewart1983} The total specific heat of a material is not a single quantity but rather the sum of several distinct contributions that reflect different degrees of freedom within the system. Separating these components makes specific heat a powerful tool for identifying phase transitions and revealing the mechanisms behind them. 

Cryogenic calorimetry performance is governed by both thermometer choice and thermal link design, each imposing distinct constraints. Superconducting thermometers deliver ultra‑high sensitivity with minimal heat capacity but operate only within narrow superconducting transition ranges.\cite{Patel2022} Thermocouples offer negligible added heat capacity, though with lower sensitivity, while metallic resistors such as platinum (Pt) are stable but loose sensitivity below 50\,K. Primary thermometers, such as those based on Coulomb blockade, rely on fundamental laws of physics to provide calibration‑free absolute temperature measurements, but they operate only within specific temperature regimes.\cite{Cole2025} In parallel, thermal links have evolved from wire‑suspended substrates with high parasitic heat capacity\cite{Stewart1983} to low‑heat‑capacity membranes, all designed to minimize background and optimize thermal conductance.\cite{Bachmann1972} Together, these advances define the trade‑offs in sensitivity, operating range, calibration, and magnetic‑field compatibility that are especially critical for micro- and nano-calorimetry.

Application‑specific calorimetric sensors designed for thin film and nanostructure samples must minimize background contributions, often requiring membranes thinner than 50\,nm.\cite{Denlinger1994,Queen2009} In some cases, excess membrane surrounding the active area is removed to further reduce thermal link, though this comes at the expense of mechanical stability.\cite{Wickey2015} Microcrystalline and microgram samples are typically studied with sub‑200\,nm membranes. For such samples, the higher background compared to thin‑film applications does not significantly hinder measurement fidelity. However, above 3\,K, the thermal grease used to mount microcrystals becomes the dominant background, diminishing the advantages of very thin nitride membranes.\cite{Tagliati2012,Schink1981} The use of such thermal agents remains an open challenge in the calorimetry community, particularly for manual mounting of microcrystals. Moreover, excessively low thermal conductance can slow calorimeter response times, underscoring the delicate balance between sensitivity, background reduction, and dynamic performance. Temperature‑modulated calorimetric methods, particularly the ac steady‑state technique,\cite{Sullivan1968} offer exceptional resolution for detecting subtle changes in heat capacity. Achieving absolute accuracy and high resolution relies on rigorous calorimeter design and a thorough understanding of the system’s thermal dynamics.

Currently available commercial systems lack the required flexibility, sensitivity, cryogenic temperature range and integration required for calorimetric studies of quantum materials. Custom platforms are required that allow precise control of thermal conductance, heater and thermometer geometry, and compatibility with various cryostats-critical when conventional heater or thermometer materials such as thermocouples and metallic thermometers loose sensitivity at low temperatures restricting operating range. Routine studies of quantum materials require ac steady‑state measurements with enhanced noise rejection and a modular mounting for rapid chip replacement. As opposed to commercial platforms which utilize slower thermal relaxation methods and are restricted to milligram-scale samples. Yet, the demand for disposable platforms remains high, as mounted crystals make chips difficult to reuse, necessitating high-throughput batch fabrication with a rapid design-to-fabrication turnaround. Integration with synchrotron x‑ray beamlines and ultra high vacuum (UHV) environments requires open geometries and specialized cryostat plugins to ensure unobstructed pathways for both the incident and scattered x-rays. Engineered for use with a wide variety of materials systems, instrument configurations and x-ray techniques, our designs span wide temperature ranges, accommodate increased x-ray flux, and employ optimized pad layouts matched to cryostat interfaces. The thermal conductance is precisely controlled through adjustments in SiN$_x$ thickness and perforations, while enabling millikelvin operation in magnetic fields using suitable heater and low‑magnetoresistance thermometers. The structures incorporate minimal addenda, ensuring high resolution measurements of sub‑microgram crystals and thin films. With these design elements in mind, the calorimeters presented here have been engineered to provide a robust platform for multimodal studies with the precision and sensitivity required for studies of phase transitions, electronic heat capacity, and structural dynamics in quantum materials, while still providing a pathway for future developments with the incorporation of electrical, mechanical or optical stimulation.   

\section{\added{Calorimeter design and fabrication}}

\begin{figure}[!htbp]
\includegraphics[width=.48\textwidth]{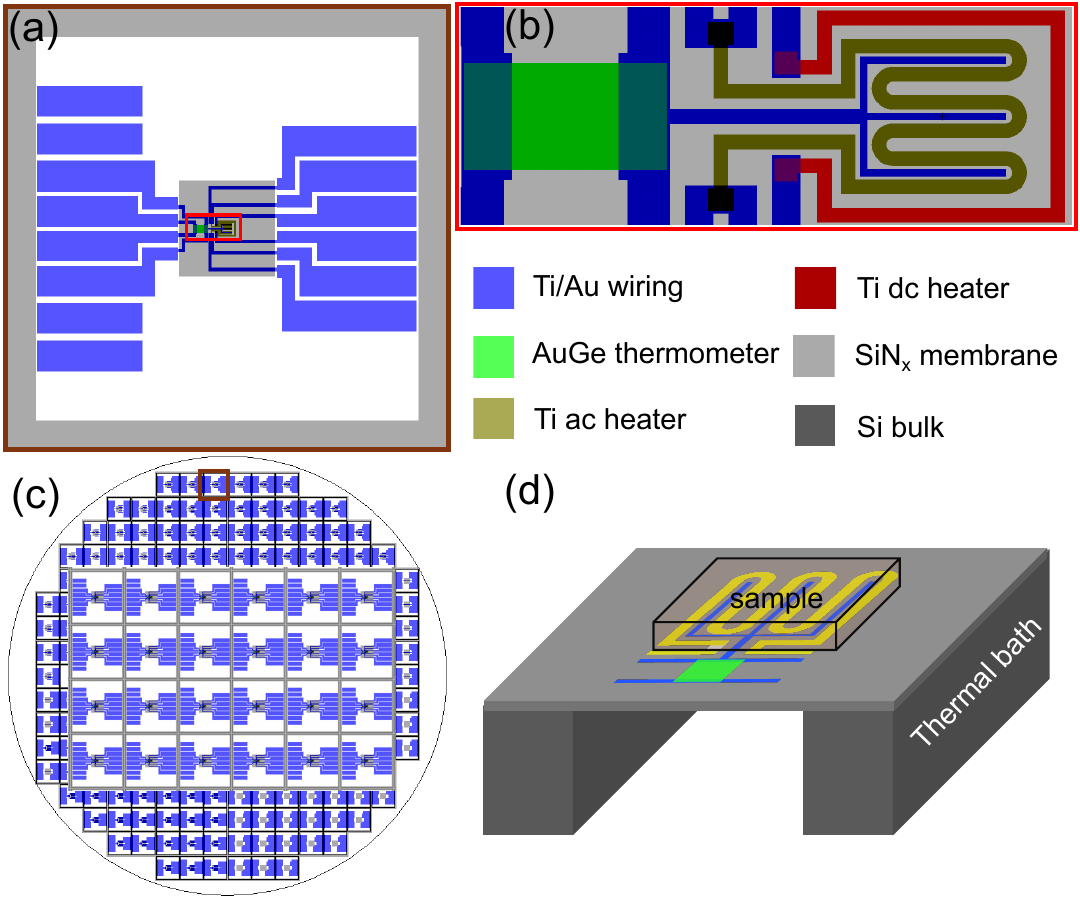}
\vspace{0.0pt} \caption{(a) Full layout of the $3.8 \,\text{mm} \times 3.8 \,\text{mm}$ chip showing the dicing lane on the chip perimeter and the square membrane in the center, $1 \,\text{mm} \times 1 \,\text{mm}$ area (gray). Bonding pads and electrical traces (blue) lead to the central active sensor area on the Si frame. (b) Central area showing color-coded layers for the GeAu thermometer, and Ti ac and dc heaters laterally arranged in the center of the membrane. A high thermal conductance Au trace connects the thermometer to the isothermal heater area in the center. (c) Array of 225 calorimetric sensor chips on a 3-inch Si wafer with enlarged view. (d) 3D cutout view of the $550 \,\text{nm}$ thick SiN$_x$ membrane acting as a weak thermal link, with the silicon frame serving as a thermal bath. The sample is typically placed on the serpentine ac resistive heater for sinusoidal Joule heating. Amplitude and phase lag of the generated thermal waves are detected at the thermometer location.}
\vspace{0.0pt} 
\label{fig:figure1}\end{figure}

Nanocalorimeters are typically built on pre-fabricated silicon nitride membranes, a process that often confines fabrication to the chip level and increases risk of membrane damage, resulting in low overall yield. To overcome this limitation, while meeting the design criteria described above, we developed a high-yield, wafer level batch-fabrication process to enable mass-production of our calorimetric sensor chip. The devices are fabricated using a 3-inch, \SI{350}{\micro\meter} thick Si wafer uniformly coated with 550\,nm thick thermally grown $\mathrm{SiN}x$ on both sides of the wafer. Figure~\ref{fig:figure1}(a) presents a top-down view of the calorimeter layout. As illustrated in Fig.~\ref{fig:figure1}(b), each cell comprises an ac heater, a thermometer, and an offset heater, all centrally positioned on the membrane. In Fig.~\ref{fig:figure1}(c), the design showcases a 3-inch silicon wafer systematically arranged with $3.8\,\mathrm{mm} \times 3.8\,\mathrm{mm}$ nanocalorimeter chips in a uniform grid. Each chip is precisely positioned with integrated dicing lanes, enabling efficient post-fabrication separation and supporting a scalable, high-throughput fabrication strategy that yields approximately 225 chips per wafer with a few design variants such as compact solid substrate thermometer/heater devices. The fabrication process includes photolithography with single or double-layer lift-off, \textit{in situ} ion-mill or RF clean with metal deposition via sputtering, reactive ion etching of the $\mathrm{SiN}x$ layer, and silicon bulk micromachining using deep reactive ion etching (DRIE). The typical fabrication layer stack and its corresponding layer thicknesses are summarized in a Table~\ref{tab:layerstack}.

\begin{table}[!htbp]
\vspace*{0.0pt}\caption{\label{tab:table4} Fabrication layer stack and thickness metrics.}
\begin{ruledtabular}
\label{tab:layerstack}
\begin{tabular}{ccc}
Index&
\makecell{Layer}&
\makecell{Layer thickness}\\
\hline
1&Ti/Au wiring&5/85\,nm\\
2&GeAu thermometer&60\,nm\\
3&Ti or Pt ac heater&70\,nm\\
4&Ti or Pt offset heater &70\,nm\\
5&Membrane perforation &550\,nm\\
6&Si bulk&\SI{350}{\micro\meter}\\
\end{tabular}
\end{ruledtabular}
\vspace*{0.0pt}
\label{table:table1}
\end{table}

The ac heater, illustrated in Fig.~\ref{fig:figure1}(b) and (d), is a serpentine-shaped titanium (Ti) or platinum resistor deposited via sputtering. It spans the central area with a trace width of \SI{10}{\micro\meter} and a thickness of 70\,nm. Designed to generate a well-defined oscillatory power, it enables a controlled, small fractional modulation of the sample temperature, typically on the order of $0.1\text{–}1\%$ across the cryogenic range. A four-point probe geometry is employed to ensure precise measurement of the power delivered to the sample, minimizing the influence of lead or contact resistance. The serpentine geometry, combined with the high electrical resistivity and lower thermal conductivity of titanium relative to gold, ensures that Joule heating is concentrated within the titanium heater region. This design minimizes parasitic heating along the gold leads before the four‑point contacts. The design minimizes extra thermal interfaces between sample and the heater improving thermal coupling to the sample. For low temperature dilution refrigeration operation, platinum is deposited as a heater material because Ti superconducts below 0.7\,K making it unsuitable as a heater at ultra low temperatures. A \SI{10}{\micro\meter}-wide trace of gold (Au), selected for its very high thermal diffusivity compared to that of the membrane, was used to connect the thermometer to the isothermal region formed by the heater and the sample. The layout is engineered so that both the offset heater and the ac heater can be deposited in a single step. 

The active sensing element of the thermometer is a precisely defined $76\,\text{µm} \times 76\,\text{µm}$ square composed of a 60\,nm thick GeAu alloy film deposited by sputtering. This micro-scale geometry ensures rapid thermal response and minimal heat capacity contribution, making it ideal for high-resolution temperature measurements. The sputtering system was configured with both RF and DC sources, utilizing RF for the GeAu alloy target and DC for Ti and Au target. The $\mathrm{Ge}_{1 - x}\mathrm{Au}_{x}$\cite{Bethoux1995} target (with $x$~=~0.17, 99.999\% purity) employed in this study was supplied by AJA International. Given the semiconducting nature of the 2-inch target, the power density was carefully maintained below $5\,\mathrm{W/cm^2}$ to prevent unwanted heating. The GeAu deposition process was initiated by striking the plasma at a pressure of 30\,mTorr, using an initial power of 55\,W with a controlled ramp-up rate of 1\,W/s to ensure a stable ignition. Once the plasma was successfully ignited, the deposition was carried out at a steady pressure of 3.5\,mTorr while maintaining a constant power of 55\,W. The GeAu sensor layer was deposited on Ti~(5\,nm)/Au~(85\,nm) wiring traces which are directly connected to the bonding pads. Following deposition, GeAu thin films were annealed at $185\,^\circ\mathrm{C}$ for one hour to stabilize their microstructure and enhance temperature-dependent resistivity. After being heated above $185\,^\circ\mathrm{C}$, all subsequent process steps were restricted to heating the wafer below $130\,^\circ\mathrm{C}$. This annealing step promotes interdiffusion between gold and germanium, resulting in low noise, high sensitivity, and reproducible resistance behavior across the full operating range of 0.1\,K to 300\,K, ensuring reliable performance for thermal sensing. The GeAu thermometer exhibited a room temperature sheet resistance of approximately $1\,\mathrm{k\Omega}$, with a dimensionless sensitivity around 1 from 300\,K down to 2\,K.

Photolithography steps, including frontside and backside wafer alignment, were carried out using MLA-150 maskless aligner equipped with a 450\,nm laser. Plasma-based dry etching of $\mathrm{SiN}x$ was employed to perform blanket etching on the wafer's backside prior to DRIE, and to create perforations in the frontside $\mathrm{SiN}x$ membranes. DRIE was performed using Plasmatherm Systems to achieve highly anisotropic bulk etching of a \SI{350}{\micro\meter} thick silicon handle layer. A \SI{7}{\micro\meter} thick layer of SPR-220 photoresist was applied as the etch mask, allowing for the simultaneous release of all device membranes and chips separation in one streamlined fabrication step. 

\section {\added{Calorimetry experimental setup with laboratory and beamline cryostats}}
\subsection{Measurement setup}

\begin{figure}[!htbp]
\includegraphics[width=.48\textwidth]{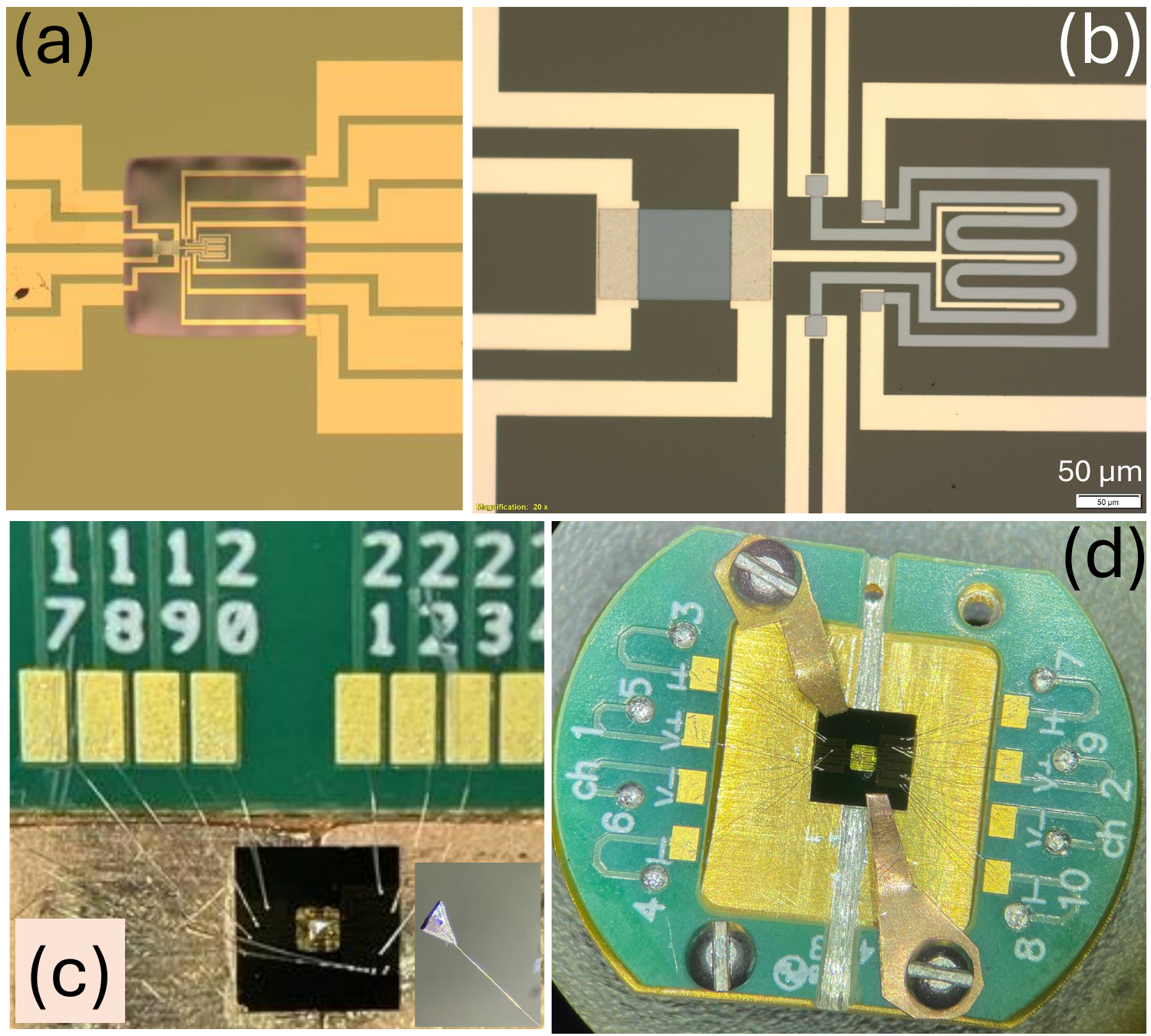}
\vspace*{0.0pt} \caption{(a) Optical microscopy of the fabricated calorimeter chip. The suspended membrane in the center of the Si frame appears transparent. (b) Optical microscope image of the central part of the calorimeter cell. (c) Calorimeter sample cell wire bonded onto the laboratory cryostat sample holder. The superconductor sample can be seen on the membrane. A vacuum channel was constructed using a pair of copper pieces. Empty‑cell and superconductor Nb sample characterizations were performed in this commercial ADR cryostat. In the bottom-right corner, a sample collected with a thin wire is displayed. (d) Dilution refrigerator sample mount from Quantum Design, made of gold‑coated silver with a pair of four‑probe channels adapted for calorimeter mounting. A pair of phosphor‑bronze clamps secured with screws hold the chip in place. N‑grease is sufficient to keep the chip fixed, but repeated thermal cycling can push it upwards.}
\vspace*{0.0pt}
\label{fig:figure2}\end{figure}

Our testing of the different design variants of calorimetry chips starts with complete device characterization in a laboratory cryostat at the Advanced Photon Source followed by variety of scientific sample testing either in x-ray beamline cryostats or in a dilution refrigerator with magnetic field capability at the Argonne National Laboratory. Figure~\ref{fig:figure2}(a) shows the optical micrograph of a fabricated calorimeter chip, while Fig.~\ref{fig:figure2}(b) highlights its central region. Micromanipulator is typically used to place small samples on the $110\,\text{µm} \times 110\,\text{µm}$ heater area on the membrane. A uniform layer of Apiezon N grease was applied to the heater surface to ensure optimal thermal coupling between the sample and the platform. The grease was carefully spread to fully cover the heater region, thereby eliminating the possibility of direct metallic conduction.  Subsequently, the sample was positioned on the membrane with precision, using a fine gold or aluminum wire for delicate handling, as illustrated in the Fig.~\ref{fig:figure2}(c). Alternatively, a thin SiO$_x$ layer can provide electrical insulation without strict step coverage or pinhole requirements, and may also be deposited as a final layer for robust isolation. The sensor chip was mounted again using Apiezon N grease to enhance the thermal contact between the silicon frame and the sample mount. A vacuum channel, as shown in Fig.~\ref{fig:figure2}(c) and Fig.~\ref{fig:figure2}(d), prevents trapped air underneath the membrane and avoids membrane breakage during pump down. Second generation calorimetry chips include membrane perforations that allows releasing air through these vacuum channels. Electrical connections were established using aluminum wire bonds between the Ti/Au contact pads on the silicon frame and the gold pads on the cryostat sample plugins. If required, the differential operation can be implemented by mounting two chips, one corresponding to the reference cell and the other to the sample cell. The complete assembly was then positioned at the 3\,K cold stage within an HPD Inc. adiabatic demagnetization refrigerator (ADR laboratory cryostat). Sixteen twisted pair wires carry ac/dc signals from 100\,mK (3\,K) to 300\,K. Outside the cryostat, DB-25 shielded cable carry signals to a custom breakout board populated with precision resistor circuits including $15\,\mathrm{k\Omega}$ and $100\,\mathrm{k\Omega}$ with a tolerance of $0.1\%$ and RJ-45 adapters. The combination of resistors and voltage source from a lock-in amplifier bias the calorimeters. Alternatively, an external low-noise current source can also be used. Three RJ-45, 568B Cat 6 cables then connect to a SynkTek MCL1-540 multichannel lock-in system.\cite{SynkTeK} A calorimeter connect-disconnect procedure was prepared and followed to avoid voltage transients and electrostatic discharge through the active on-chip components.   

\subsection{Multimodal cryostat plugins}

\begin{figure}[!htbp]
\includegraphics[width=.48\textwidth]{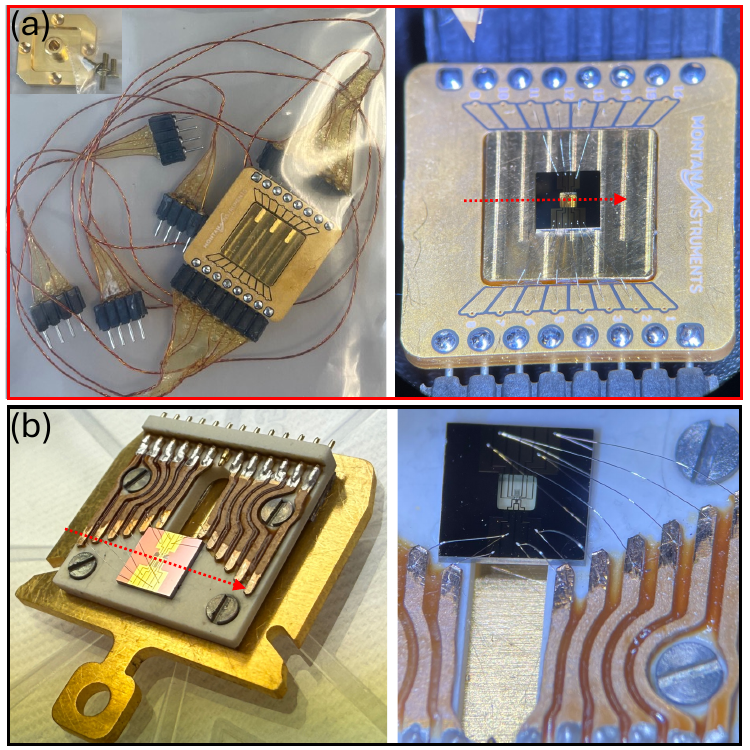}
\vspace*{0.0pt} \caption{Beamline cryostat plugins and calorimetry with multimodal geometry: (a) Packaged calorimeter chip connected to gold‑plated wire bonding pads on a commercial CC16 cryostat plugin with eight twisted‑pair wires and four matching male connectors (16 pins) for plugging into the adapter interface at the 4\,K stage of the Montana Instruments Cryostation S100 cryostat. The feed‑through adapter provides an interface to a room‑temperature MDR cable for electrical connection to lock‑in electronics. The sample holder can be mounted in either a horizontal or vertical orientation using a screw‑in base adapter support. (b) Ferrovac EC13 UHV sample holder with 13 spring‑loaded probe contacts for electrical connections, adapted with CuBe$_2$ gold‑plated wire bonding traces attached using UHV epoxy. The white ceramic base plate is modular and can be removed from the CuBe$_2$ gold‑plated base.}
\vspace*{0.0pt} 
\label{fig:figure3}\end{figure}

The sample mount used in Bluefors cryogen-free dilution refrigerator is shown Fig.~\ref{fig:figure2}(d). The chip was secured with custom phosphor-bronze spring clamps, as illustrated in Fig.~\ref{fig:figure2}(d).  This setup was used to perform aluminum superconducting phase transition measurements under different magnetic fields as described in Section~\ref{sec:results}. The cryostat plugins for multimodal setups combining x-ray scattering and heat capacity measurements are illustrated in Fig.~\ref{fig:figure3}. At 6-ID-C, which is dedicated to hard x-ray scattering, a single calorimeter cell wire-bonded to a CryoChip16 (CC16) cryostat plugin from Montana Instruments is shown in Fig.~\ref{fig:figure3}(a). The packaged calorimetry plugin shown provides unobstructed passage of the incident x‑ray beam to the sample and the scattered beam away from it. Independent measurements using a cryostat at the 6-ID-C beamline verified ac steady‑state calorimetry operation and wide‑range offset scanning through the on‑chip offset heater, accomplished without altering the cryostat temperature. In the case of the UHV diffractometer at 29-ID which features a Ferrovac EC13 sample receiver that is connected to the cryostat via a copper braid, semi-permanent low temperature UHV epoxy Torr Seal was used to mount the chip (Fig.~\ref{fig:figure3}(b)). After chip packaging, the sample holder is typically placed on the transfer arm and then engaged and locked in the cryo-receiver of the beamline cryostat. The custom cryostat plugin packaged with the calorimeter chip was cooled down and tested for electrical functionality of the heater and thermometer. The temperature dependence confirmed that the spring-loaded electrical interface works reliably, along with completing the thermometer calibration (data not shown).  The cutout in the ceramic base plate can be used to hold a pre-calibrated in-house GeAu or commercial silicon diode thermometer for initial calibration, if a new chip is not externally pre-calibrated in a laboratory cryostat.   

\subsection{GeAu thermometer calibration}

\begin{figure}[!htbp]
\includegraphics[width=.48\textwidth]{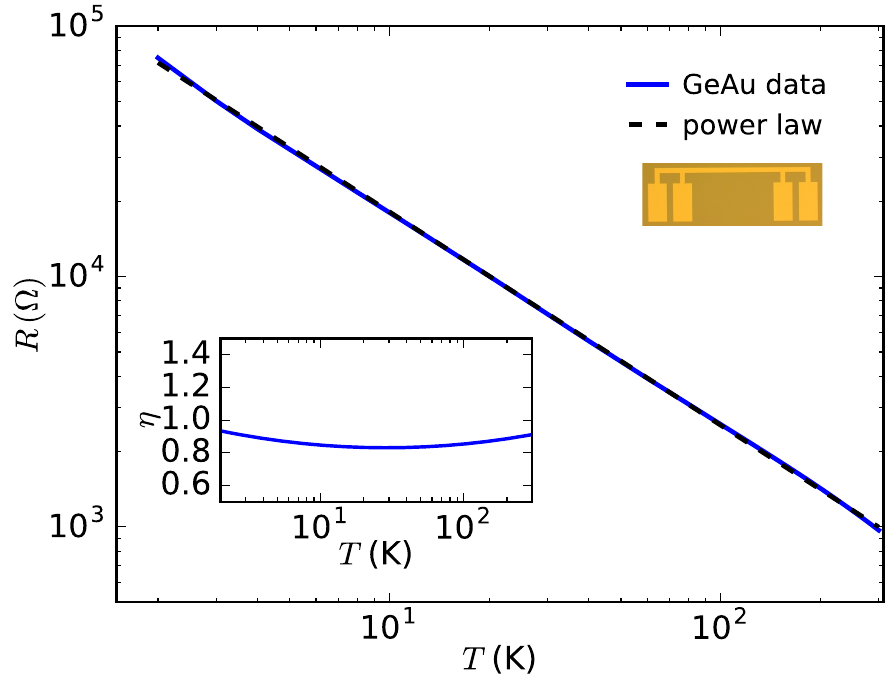}
\vspace*{0.0pt} \caption{Measured (blue) temperature dependence of the one-square GeAu thermometer with a single-exponent power-law fit (dashed). The inset shows the dimensionless sensitivity across the full temperature range, obtained using parameters generated by fitting the analytic expression to the calibration data, as discussed in the main text. The inset also shows an image of the four-probe control device for GeAu, consisting of 25 squares for tracking effect on GeAu as wafer progresses through complete fabrication run. (measured data not shown).}
\vspace*{0.0pt} 
\label{fig:figure4}\end{figure}

Calibration of GeAu thermometers typically begins by referencing a well-characterized primary thermometer, such as a calibrated Cernox and $\mathrm{RuO}_{2}$, over the desired temperature range. The GeAu sensor is placed in thermal contact with the reference thermometer, and its resistance is recorded as the temperature is varied in a controlled manner. This yields a dataset of resistance versus temperature as shown in Fig.~\ref{fig:figure4}, which serves as basis for generating a calibration curve. The GeAu resistance bridge (inset), consisting of 25 squares, and a separate device comprising a single GeAu square with Ti/Au contacts on a solid substrate, were also independently tested (data not shown) and compared to ensure consistent and robust performance under aging and thermal cycling. The calibration curve is generally consistent across different devices and can serve as an initial check by normalizing resistance against the available calibration data. For temperature intervals spanning only a few tens of kelvin, calibration parameters can often be captured with a single power exponent. For wide temperature range, we employed a calibration approach for GeAu thermometry that expresses temperature explicitly as a function of resistance. Specifically, we express the logarithm of temperature as a polynomial expansion in the logarithm of resistance:
\begin{equation}
\ln T = \sum_{k=0}^{3} b_k \, (\ln R)^k
\label{eq:RTrelation}
\end{equation}
This formulation yields a closed-form expression for temperature and eliminates the need for numerical inversion of log-polynomial fits. From this model, the sensitivity function $\eta(R)$, defined as the inverse of the logarithmic derivative of temperature with respect to resistance, becomes:
\begin{equation}
\eta(R) = \left| \frac{d \ln T}{d \ln R} \right|^{-1} 
= \frac{1}{\sum_{k=1}^{3} k \, b_{k} \, (\ln R)^{\,k-1}}
\end{equation}

This approach provides smooth analytical control over both $T(R)$ and $\eta(R)$, making it ideal for high-precision calibration across a broad temperature range. The $R–T$ dependence of doped Germanium thermometers is frequently fitted using Equation~\ref{eq:RTrelation}.\cite{Pobell2007} 

\section {\added{Calorimetric measurements and analysis}}

AC steady-state calorimetric response is governed by thermal links between the sample, heater, thermometer, and substrate. Models for conventional \cite{Sullivan1968} and substrate-based calorimeters\cite{Tagliati2011,Velichkov1992,Gmelin1997} address measurement accuracy, including techniques that assess sample-substrate coupling via phase ($\phi$) analysis and low frequency ac non-adiabatic conditions. Dynamic calorimeters with internal relaxation are described using complex heat capacity, thermal analogue of an electrical RC circuit, with real (in-phase) and imaginary (out-of-phase) components.\cite{Ema1997}

\subsection{AC steady state frequency dependence: reference and sample cell}

In ac steady-state calorimetry,\cite{Sullivan1968} when the ac heater is driven by an electric current \(I(t) = I_0 \cos \omega t\), the power dissipated in the heater resistor \(R_h\) is \(P(t) = P_0 (1 + \cos 2\omega t)\), where \(P_0 = I_0^2 R_h / 2\). The resulting temperature response of the cell can be expressed as \(T(t) = T_b + T_{dc} + T_{ac}\cos(2\omega t + \phi)\). Here, \(T_b\) is the bath temperature, while \(T_{dc} = P_0/K_e\) represents the steady temperature rise above the bath due to the average heating power, with \(K_e\) being the thermal conductance that governs heat flow from the sample to the bath. The oscillatory term \(T_{ac}\cos(2\omega t + \phi)\) describes the periodic modulation of the temperature at twice the driving frequency, where \(T_{ac}\) is the amplitude of the oscillations and is directly related to the heat capacity of the sample. The parameter \(\phi\) denotes the phase lag, reflecting the delay between the oscillating heating power and the temperature response. The heat capacity and thermal conductance in terms of measured parameters ($T_{ac},\,\phi,\,\omega,\,P_{0}$) are then given by \begin{align}
C &= \frac{P_0}{2 \omega T_{ac}} \sin \phi \\
K &= \frac{P_0}{T_{ac}} \cos \phi
\end{align}

\begin{figure}[!htbp]
\includegraphics[width=0.48\textwidth]{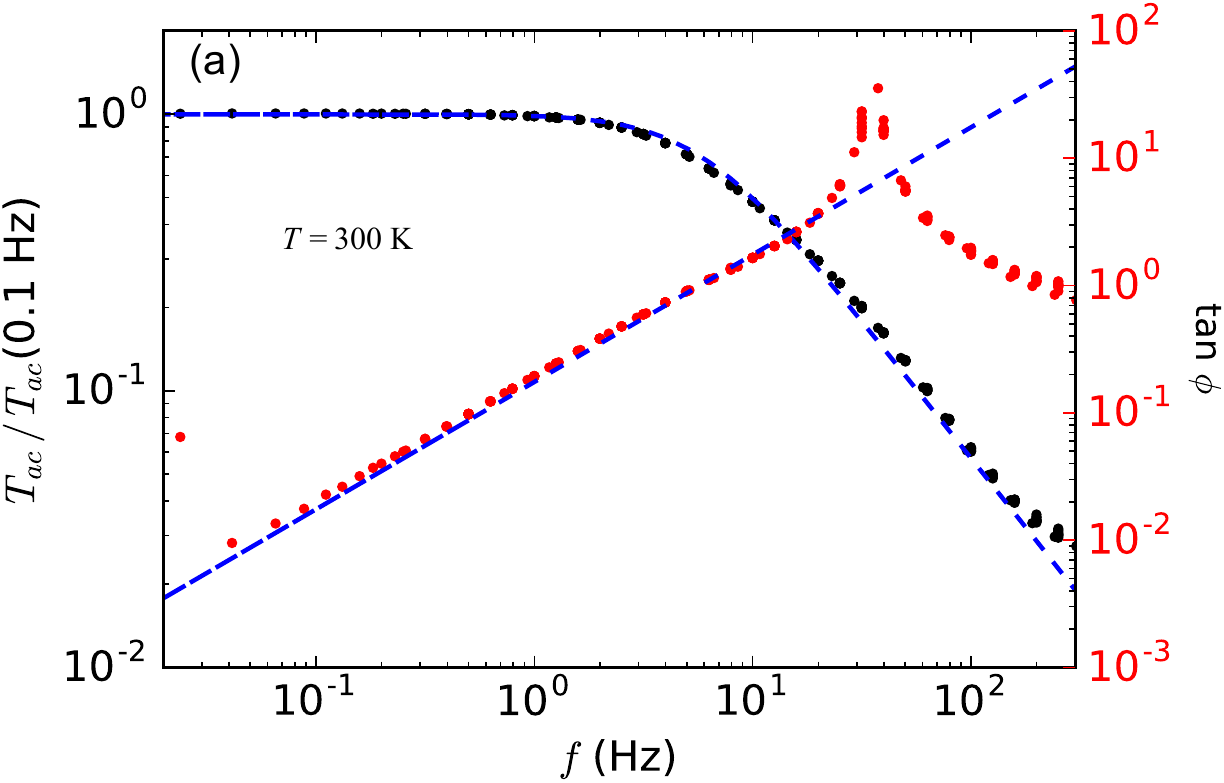}
\includegraphics[width=0.48\textwidth]{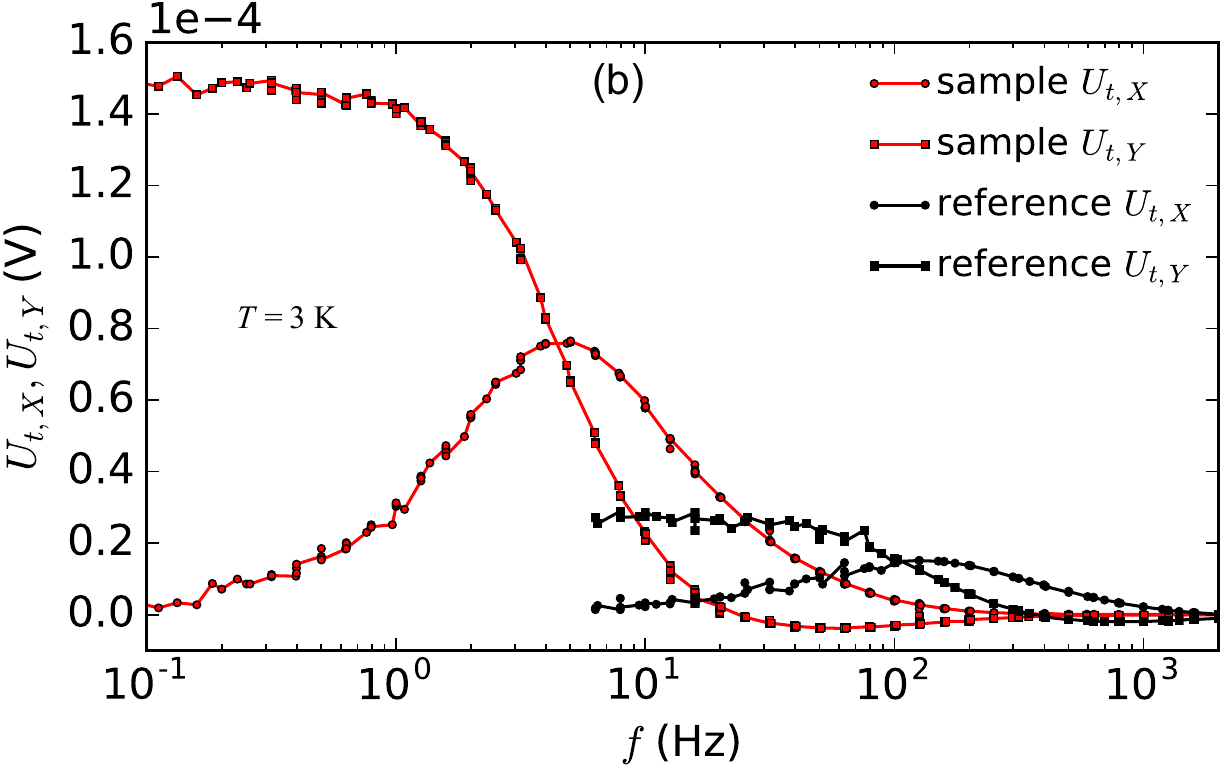}
\vspace*{0.0pt} \caption{(a) Measured frequency response of the calorimeter cell at 300\,K showing normalized amplitude $T_{\mathrm{ac}}$ and phase lag $\phi$ expressed as $\tan \phi$. Both dashed lines are corresponding fits from $\tan \phi \propto \omega \tau_e$ and $T_{\mathrm{ac}} \propto 1/\sqrt{1+\omega \tau_e}$ with the same 
value for parameter $\tau_e$. The obtained system time constant matches very well with the one measured using the dc relaxation method. (b) Measured frequency dependence of the thermometer ac voltages in-phase ($U_{tX}$) and out-of-phase ($U_{tY}$) lock-in components of the Nb sample and reference cells at 3\,K. The typical operating frequency point, defined by $\tan \phi = 1$, is easily determined where both components are equal in magnitude. For the empty cell, $\tan \phi = 1$ occurs at $f_{\mathrm{op}} \approx 10\,\text{Hz}$ at 300\,K, shifting to $\sim 100\,\text{Hz}$ at 3\,K, showing the operating frequency range of the empty cell. For the Nb sample cell at 3\,K, this point shifts to lower frequency, highlighting the need to determine the working frequency for each measured sample.}
\vspace*{0.0pt} 
\label{fig:figure5}\end{figure}

These relations are practically useful under the assumption that the calorimetric system operates in a quasi-adiabatic regime, typically achieved by selecting an appropriate operating frequency such that \( \omega \tau_i \ll 1 \). Here, \(\tau_i\) is the internal relaxation time between the sample and the calorimetric cell. In practice, a frequency dependence of \( T_{\text{ac}} \) and \( \tan(\phi) \) provides a good range of operating frequencies where both accuracy and resolution can be optimized. One such plot is shown in Fig.~\ref{fig:figure5}(a). This can also be readily achieved by examining the raw plot (Fig.~\ref{fig:figure5}(b)) of frequency dependence of \( X \) in‑phase component and \( Y \) out‑of‑phase component. At the point where, \( \tan(\phi) = 1 \) the phase angle \( \phi = 45^\circ \), equivalently \( \omega \tau_e = 1 \). Here, \(\tau_e = C / K_e\) is the external relaxation time between the sample and the thermal bath. The transfer function of our calorimetric system also shows a clear plateau where the product $\omega$\( T_{\text{ac}} \) is constant (not shown). Our thermometer follows a power-law model \( R(T) = R_0 \left( \frac{T}{T_0} \right)^\eta \). If the temperature oscillations \( T_{\text{ac}} \ll T_0 \), the measured thermometer ac voltage, according to the leading order of a Taylor expansion, becomes
\begin{equation}
U_{\text{ac}}(t) \approx I_{\text{dc}} R_0 \eta \frac{T_{\text{ac}}}{T_0} \cos(2\omega t + \phi)
\end{equation} The lock-in amplifier extracts the amplitude \( U_{\text{ac}} \) which results in
\begin{equation}
T_{\text{ac}} = \frac{U_{\text{ac}}}{U_{\text{dc}}} \cdot \frac{T_0}{\eta}
\end{equation}
and phase \( \phi \) of the thermometer signal at the temperature modulation frequency \( 2\omega \).

\subsection{Empty cell characterizations}

\begin{figure}[!htbp]
\includegraphics[width=.48\textwidth]{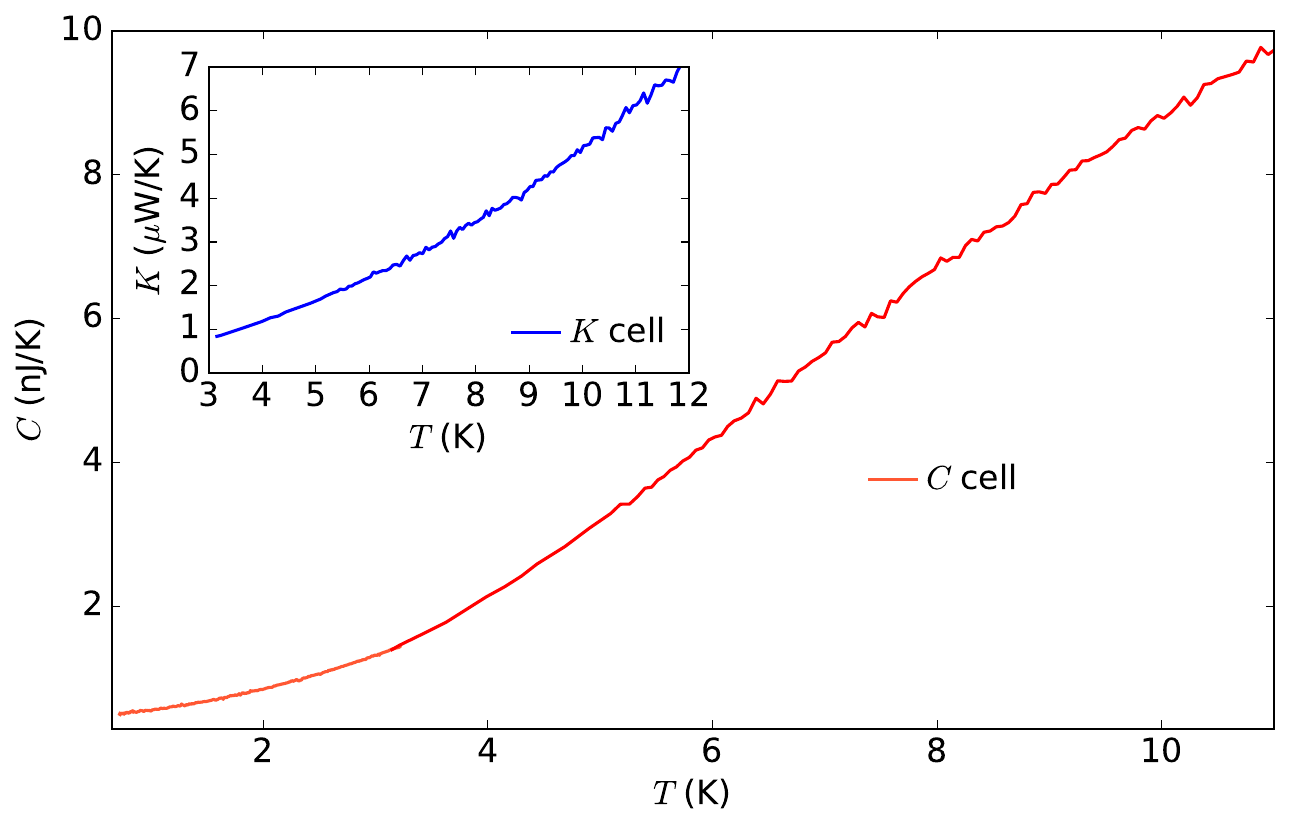}
\vspace*{0.0pt} \caption{AC steady-state characterization of the calorimeter cell. Measured heat capacity of the empty cell at low temperatures relevant for superconducting phase transitions providing addenda heat capacity. The inset shows measured thermal conductance of the empty cell.}
\vspace*{0.0pt} 
\label{fig:figure6}\end{figure} 

The empty cell characterization temperature range is selected based on the samples under evaluation. An empty cell nanocalorimeter measurement is the essential baseline step in high-sensitivity calorimetry, because it defines the thermal properties of the calorimeter platform itself before any sample is mounted. In this configuration, the cell’s effective heat capacity ($C_{\text{cell}}$) is determined by applying a sinusoidal heating power and measuring the resulting temperature oscillation amplitude and phase lag, which reveals how much energy the membrane, heater, thermometer, and grease layers store. At the same time, the thermal conductance ($K_{\text{cell}}$) of the cell is extracted from the out-of-phase component of the response, describing how efficiently the calorimeter exchanges heat with the thermal bath. Together, $C_{\text{cell}}$ and $K_{\text{cell}}$ define the external time constant of the system, set the limits for quasi-adiabatic operation, and provide the addenda background that must be subtracted from subsequent sample measurements to isolate true sample-specific heat capacity. The measured heat capacity of the empty cell at low temperatures below 12\,K is shown in Fig.~\ref{fig:figure6}. The measured heat capacity of the cell ranges from 380\,pJ/K at 0.75\,K to 9.7\,nJ/K at 11\,K above the phase transition temperatures of Nb and Al. While the current device already demonstrates excellent performance, reducing the $\mathrm{SiN}_{x}$ thickness to one‑third would further lower the cell’s background, enabling smaller sample masses and removing background as a limitation for high accuracy and resolution. This change can be easily accomplished as it does not require any change in design or fabrication process flow. However, the trade‑off between lower empty‑cell addenda and higher thermal conductance necessitates the use of two separate cells, each optimized for its respective requirement. The simultaneously obtained thermal conductivity of the empty calorimetry cell is shown in the inset of Fig.~\ref{fig:figure6}. The characteristic time of the cell $\tau_{\text{cell}}$ at 3\,K obtained with the ac steady-state measurements agrees fairly well with the thermal relaxation method. Though, the operating frequency plays a critical role in determination of $K_{\text{cell}}$ requiring \( \tau_i \ll 1/\omega \). By increasing the thermal conductance by an order of magnitude relative to a 200\,nm membrane cell, the system achieves a lower base temperature for a given x-ray beam power, \(P_{\text{beam}}\), as expected from \(\Delta T = P_{\text{beam}}/K\). For a cryostat operating at a base temperature of \(10\,\text{K}\), a temperature increase of \(5\,\text{K}\) and a thermal link of \SI{7}{\micro\W/K} requires an absorbed beam power of approximately \SI{35}{\micro\W}. In practice, additional beam-induced heating can be mitigated using attenuators or other flux‑reduction methods to keep the absorbed power below a specified level. 

\section {\added{Calorimetric phase transition study}}\label{sec:results} 
The low-temperature performance of our calorimeters is demonstrated through electronic specific heat measurements on two high-purity, small-scale samples of elemental superconductors, Nb and Al. These materials were selected for their markedly different intrinsic superconducting properties. The results are benchmarked against bulk samples, with masses differing by nearly seven orders of magnitude.     

\subsection{Phase transition of Nb superconductor}

\begin{figure}[!htbp]
\includegraphics[width=0.48\linewidth]{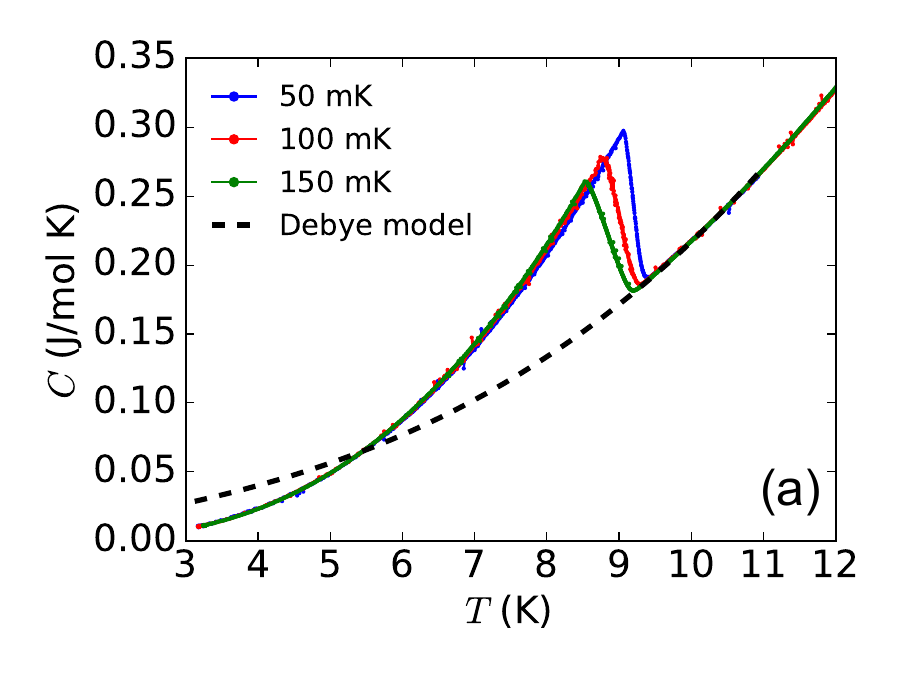}
\includegraphics[width=0.48\linewidth]{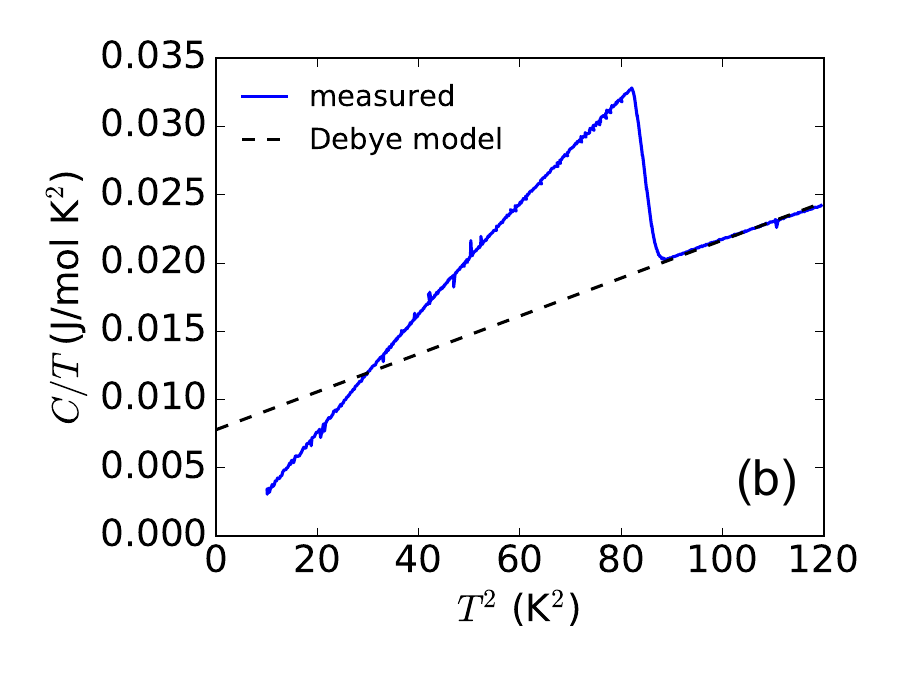}
\includegraphics[width=0.48\linewidth]{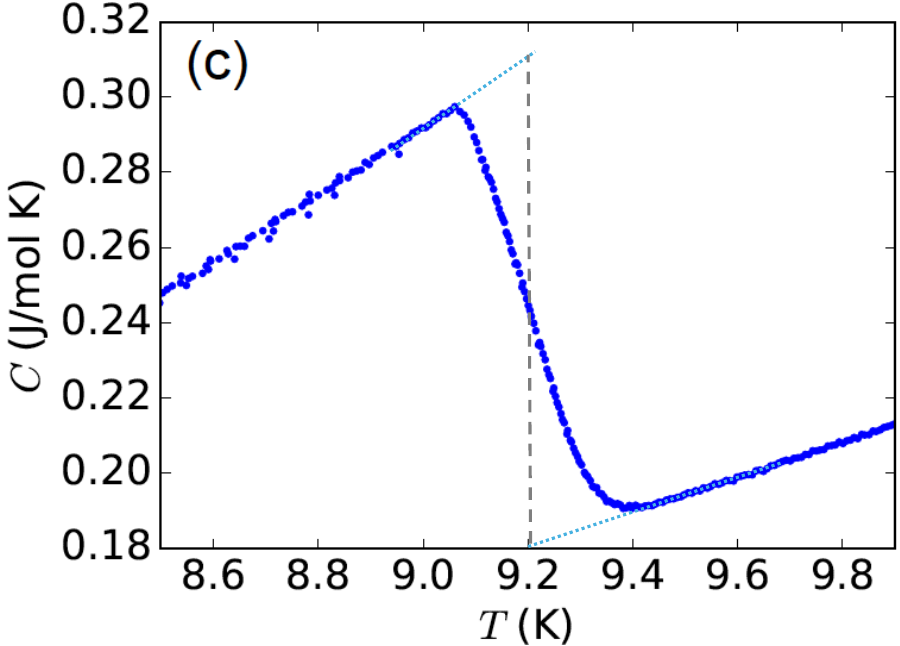}
\includegraphics[width=0.48\linewidth]{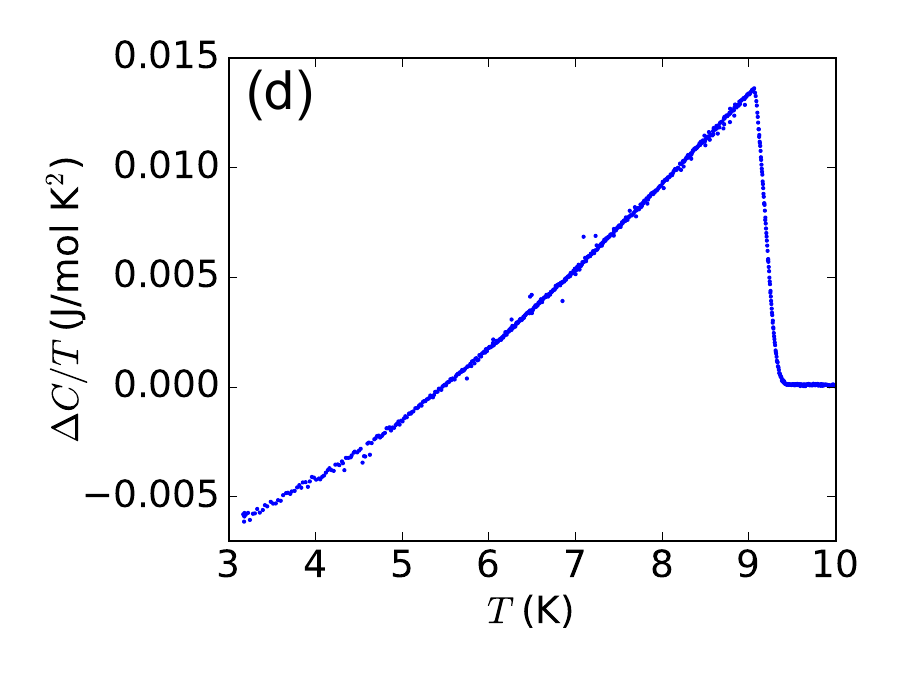}

\vspace*{0.0pt} \caption{(a) Temperature dependence of the low-temperature specific heat $C(T)$ of a Nb sample for various amplitudes of the temperature modulation. 
(b) Plot of $C/T$ versus $T^2$, showing the linear relation used to extract the electronic and phononic contributions. 
(c) Specific heat in the narrow region around $T_c$ for $T_{\mathrm{ac}} = 50\,\mathrm{mK}$, used to determine $\Delta C$. The straight lines are extrapolations drawn to the midpoint of $T_c$. The apparent broadening of the transition is due to higher ac heater drive currents. 
(d) Specific-heat difference $(C_s - C_n)/T$ after subtraction of the normal-state background. 
}
\vspace*{0.0pt} 
\label{fig:figure7}\end{figure}

The sample used for this study was manually cut from a \SI{127}{\micro\meter} thick Nb foil from Alfa-Aesar with a purity of 99.97\%. The heat capacity of the sample was measured as a function of temperature, as shown in Fig.~\ref{fig:figure7}(a). A clear jump in heat capacity is resolved at the superconducting transition temperature $9.2\,\text{K}$ indicating onset of bulk superconductivity. In the normal state, the background contribution from lattice vibrations is clearly evident and increases rapidly as the temperature rises. The contributions from the addenda and mounting grease, estimated to be less than $1\%$ of the total heat capacity at $9.2\,\text{K}$. The heat capacity contribution of the membrane and grease addenda was determined in an independent measurement run. The specific heat of the Nb sample is presented in the Fig.~\ref{fig:figure7}(a) for different values of $T_{\mathrm{ac}}$. At higher drive amplitudes, the transition becomes broadened because the larger temperature modulation averages the response over a wider range, reducing the sharpness of the discontinuity at $T_c$. The specific heat jump $\Delta C \approx 130\,\mathrm{mJ\,mol^{-1}K^{-1}}$ was determined using side extrapolations across the transition, as illustrated in Fig.~\ref{fig:figure7}(c), and the values obtained are consistent with those reported for Nb crystals of much larger mass up to ($\sim 50\,\text{g}$) in the literature.\cite{Novotny1975,Leupold1964,Carbotte1990} To place the data on an absolute scale in Fig.~\ref{fig:figure7}, the measured heat capacity was normalized to the reported value at $T_c$ for Nb.\cite{Novotny1975,Leupold1964} The curve plotted as $C/T$ versus $T^2$ linearizes the data, as shown in Fig.~\ref{fig:figure7}(b), and provides a straightforward way to separate the electronic and lattice contributions to the heat capacity. In the normal state, the data indeed follow a linear trend, with the intercept at $T^2 = 0$ yielding the Sommerfeld coefficient $\gamma$ and the slope reflecting the phonon background. Below $T_c$, the deviation from linearity highlights the loss of electronic heat capacity due to superconductivity, while any residual offset indicates nonsuperconducting contributions or incomplete background subtraction. To analyze the results, the scaled curve was then extrapolated to low temperatures and fitted to the Debye-Sommerfeld relation for the normal state,
\begin{equation} C_{n} (T) = \gamma T + \beta T^{3}, \quad \text{where} \quad \beta = \frac{12}{5}\pi^{4}\frac{R}{\Theta_D^{3}}, \end{equation}

$\Theta_D$ represents the Debye temperature associated with lattice vibrations, $R$ is the gas constant, and $\gamma$ is a coefficient proportional to the electronic density of states at the Fermi surface. In practice, the normal state specific heat $C_{n}(T)$ is obtained by applying magnetic field to suppress the superconductivity, but this cannot be done here since the ADR cryostat has no sample-field magnet. In Fig.~\ref{fig:figure7}(d), the background heat capacity $C_{n}$ from the Debye model was subtracted, and the superconducting contribution was plotted as
\begin{equation}
\frac{\Delta C}{T} = \frac{C_{s} - C_{n}}{T}
\end{equation}
versus temperature. Subtracting the normal-state background provides an accurate probe of the superconducting contribution and enables meaningful comparison. The superconducting-state heat capacity also follows the expected temperature dependence of $\Delta C/T$ reported for large Nb samples.\cite{Novotny1975,Leupold1964} Minor deviations can be attributed to the effects of annealing on superconducting properties, such as defects and dislocations. From the scaling factor, the sample mass was estimated to be approximately \SI{142}{\micro\gram}, in close agreement with the measured volume of the regular-shaped specimen and the independent mass determination of \SI{147}{\micro\gram} obtained via optical microscopy. In a separate set of measurements a \SI{65}{\micro\gram} Nb sample cut from the same foil was measured. The measured specific heat also revealed a clear phase transition. Here, we estimate the sample volume from microscopy images.

\subsection{Phase transition of Al superconductor in a magnetic field}

\begin{figure}[!htbp]
\includegraphics[width=0.48\textwidth]{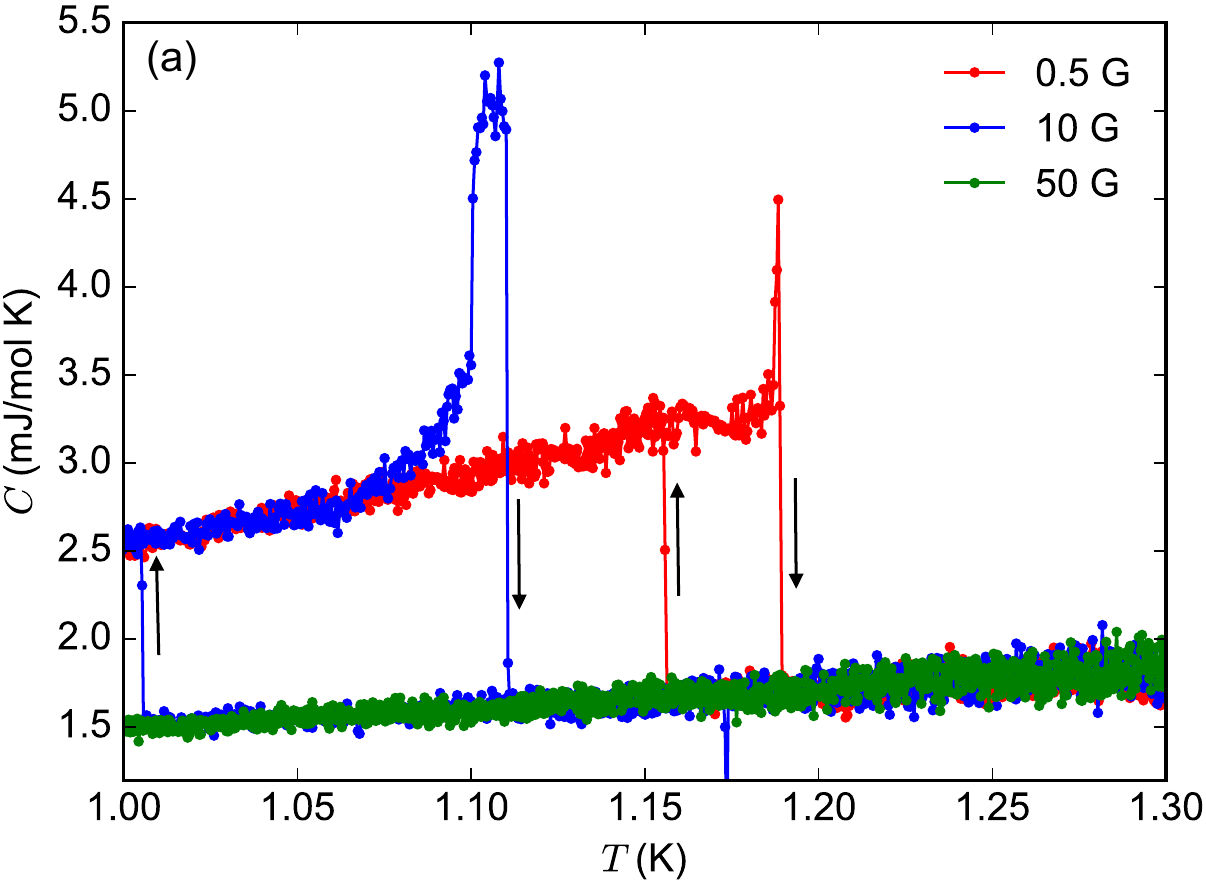}
\includegraphics[width=0.48\textwidth]{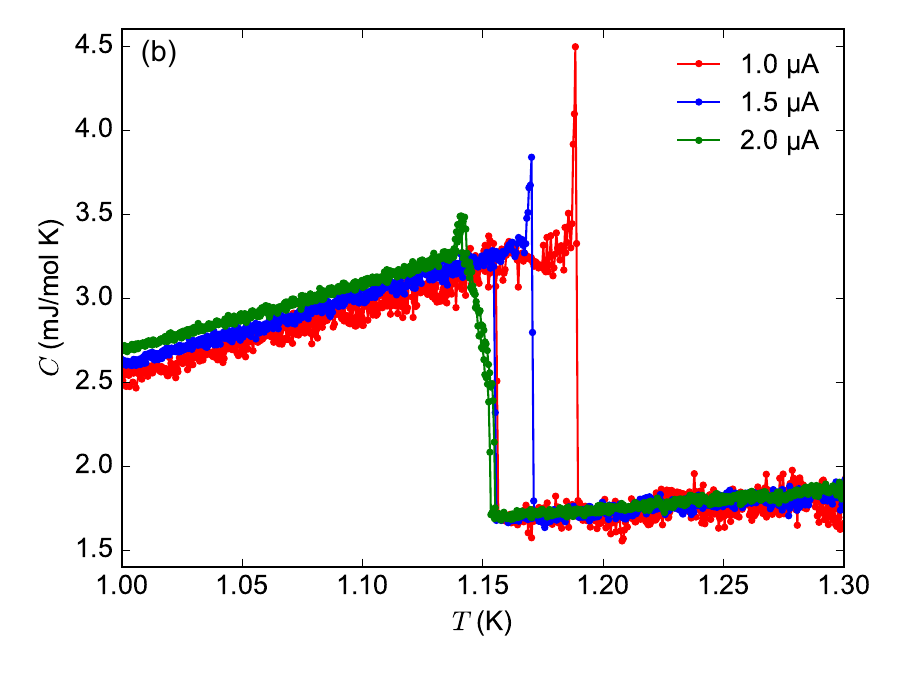}
\vspace*{0.0pt} \caption{(a) Specific heat of pure Al measured for different magnetic field strengths using ac steady-state operational parameters $T_{\mathrm{ac}} = 2.5\,\mathrm{mK}$, $f_{op}$ = 7\,Hz, $I_0$ = 1.0\,\text{µ}A. (b) The low-temperature specific heat $C(T)$ of an Al sample as a function of the applied ac heater current in a 0.5\,G magnetic field. As the current increases, both the hysteresis width and the latent-heat peak decrease in magnitude, and at sufficiently high currents they vanish.}
\vspace*{0.0pt} 
\label{fig:figure8}\end{figure}

A \SI{4}{\micro\gram} aluminum sample investigated in this high-resolution ac calorimetry measurements was manually cut from a \SI{130}{\micro\meter} thick Al foil (Alfa-Aesar, 99.9995\% purity). To avoid introducing mass uncertainty error, the measured heat‑capacity in Fig.~\ref{fig:figure8} was scaled to the reported bulk specific‑heat value at the critical temperature in the literature.\cite{Phillips1959} The heat capacity under an applied magnetic field was measured as a function of temperature using a Bluefors dilution refrigerator, as shown in Fig.~\ref{fig:figure8}(a). A pronounced jump in heat capacity is observed at the superconducting transition temperature of 1.18\,K, signifying the emergence of the superconducting phase. A linear background in the normal state indicates a weak lattice term that grows slightly near $T_{c}$, unlike the strong $T^{3}$ dependence characteristic of Nb. The measured $\Delta C$ is in close agreement with the values obtained in measurement on bulk samples.\cite{Phillips1959,Rorer1963} As shown in Fig.~\ref{fig:figure8}(a), a narrow spike is superimposed onto the step in the specific heat. Furthermore, the transition is hysteretic as indicated by the arrows. With increasing field, the spike grows in widths and height, and the hysteresis increases. This behavior is a signature of the first-order nature of the superconducting transition of a type-I superconductor, such as Al, in a magnetic field. Similar behavior has been well documented in adiabatic calorimetry on a \SI{47}{\gram} bulk sample.\cite{Rorer1963} Figure~\ref{fig:figure8}(b) shows the evolution of the superconducting transition with increasing ac heater current, that is, with increasing oscillating temperature $T_{\mathrm{ac}}$, measured while increasing the sample temperature. The transition shifts to lower temperatures while at the same time the spike is reduced and starts to broaden. The temperature shift is expected as the effective dc component of the heater power induces a temperature rise of \(\Delta T = P_{0} / K\). The apparent suppression of the first-order signature is caused by the smearing of the transition with increasing $T_{\mathrm{ac}}$, and reflects the observation that it is generally difficult to detect the latent heat in an ac specific-heat measurement.\cite{Bouquet1998,Hudl2014} Nevertheless, our ac calorimeter can operate with ac temperature amplitudes that are sufficiently small to detect first-order phase transitions, while still affording a sufficient signal-to-noise ratio.    
 
\section{\added{conclusions}}

We have developed advanced, high‑resolution, high-throughput calorimetric sample platforms tailored for thermodynamic investigations across a broad temperature range, cryogenic to room temperature, and engineered for compatibility with synchrotron x‑rays. Our tested versatile platforms enables seamless integration across diverse cryogenic systems, including beamline cryostats optimized for multimodal geometries and laboratory‑based dilution refrigerator cryostats with magnetic field capabilities. Moreover, the same chip with a mounted sample can be removed and re‑mounted across different platforms via wire bonding, providing unique flexibility for multimodal and cross‑platform studies. The calorimeter achieves exceptional energy resolution, capable of detecting latent heat peaks on the few-pJ scale, while maintaining high temperature sensitivity. The intrinsic background heat capacity of our devices remains below 0.4\,nJ/K, allowing high-resolution characterization of small \text{µg}‑scale samples. Moreover, our approach supports further reductions in background heat capacity via thin nitride starting wafers and facilitates a compact thermometer-heater assembly, suggesting that the ultimate performance limits are awaiting experimental validation.  

This capability opens the door to probing subtle phase transitions and complex thermodynamic phenomena with exceptional resolution and is particularly well suited for multimodal studies of quantum materials where the material properties evolve dramatically with small changes in temperature. Our multimodal calorimetric platforms thus emerge as powerful instruments for advancing cryogenic thermodynamics. Within this framework, even smaller energy exchanges and finer temperature variations can be detected, while simultaneously revealing the intrinsic energy landscape, and structural transformations that govern these quantum and condensed matter systems.

\begin{acknowledgements}
{This research used resources of the Advanced Photon Source and the Center for Nanoscale Materials, U.S. Department of Energy (DOE) Office of Science user facilities at Argonne National Laboratory, and is based on work supported by the U.S. DOE Office of Science-Basic Energy Sciences, under Contract No. DE-AC02-06CH11357. The authors would like to thank A. Rydh for discussions on their nanocalorimeter device fabrication particularly experience with GeAu. We also acknowledge R. Divan and S. Miller for helpful discussions on cleanroom microfabrication.}
\end{acknowledgements}

\section*{AUTHOR DECLARATIONS}

\subsection*{Conflict of Interest}
The authors have no conflicts to disclose.

\subsection*{Data Availability Statement}
The data that support the findings of this study are available from the corresponding author upon reasonable request.

\section*{References}

\renewcommand{\bibsection}{}


\begin{thebibliography}{99}

\bibitem{Stewart1983}
G. R. Stewart, 
``Measurement of low-temperature specific heat,''
\textit{Rev. Sci. Instrum.} \textbf{54}, 1 (1983).

\bibitem{Willa2017}
K. Willa, Z. Diao, D. Campanini, U. Welp, R. Divan, M. Hudl, Z. Islam, W.-K. Kwok, and A. Rydh,
``Nanocalorimeter platform for \textit{in situ} specific heat measurements and x-ray diffraction at low temperature,''
\textit{Rev. Sci. Instrum.} \textbf{88}(12), 125108 (2017).

\bibitem{schick2016}
Schick, C. and Mathot, V. (Eds.).
``Fast Scanning Calorimetry,''
Cham: Springer, (2016).

\bibitem{Russell1985}
T. P. Russell and J. T. Koberstein,
``Simultaneous differential scanning calorimetry and small-angle x-ray scattering,''
\textit{J. Polym. Sci., Polym. Phys. Ed.} \textbf{23}, 1109-1115 (1985).

\bibitem{Clout2016}
A. Clout, A. B. M. Buanz, T. J. Prior, C. Reinhard, Y. Wu, D. O’Hare,
G. R. Williams, and S. Gaisford,
``Simultaneous differential scanning calorimetry-synchrotron x-ray powder diffraction: A powerful technique for physical form characterization in pharmaceutical materials,''
\textit{Anal. Chem.} \textbf{88}, 10111 (2016).

\bibitem{Lexa2003}
D. Lexa and A. J. Kropf,
``The beam-heating effect in simultaneous differential scanning calorimetry/synchrotron powder X-ray diffraction,''
\textit{Thermochim. Acta} \textbf{401}, 239 (2003).

\bibitem{Baeten2015}
D. Baeten, V. B. F. Mathot, T. F. J. Pijpers, O. Verkinderen, G. Portale,
P. Van Puyvelde, and B. Goderis,
``Simultaneous synchrotron WAXD and fast scanning (Chip) calorimetry: On the (Isothermal) crystallization of HDPE and PA11 at high supercoolings and cooling rates up to 200 °C/s,''
\textit{Macromol. Rapid Commun.} \textbf{36}, 1184 (2015).

\bibitem{Wurm2009}
A. Wurm, A. A. Minakov, and C. Schick,
``Combining X-ray scattering with dielectric and calorimetric experiments for monitoring polymer crystallization,''
\textit{Eur. Polym. J.} \textbf{45}, 3280 (2009).

\bibitem{Bras1995}
W. Bras, G. E. Derbyshire, A. Devine, S. M. Clark, J. Cooke,
B. E. Komanschek, and A. J. Ryan,
``The combination of thermal analysis and time-resolved x-ray techniques: A powerful method for materials characterization,''
\textit{J. Appl. Crystallogr.} \textbf{28}, 26-32 (1995).

\bibitem{Xiao2013}
K. Xiao, J. M. Gregoire, P. J. McCluskey, D. Dale, and J. J. Vlassak,
``Scanning AC nanocalorimetry combined with \textit{in situ} x-ray diffraction,''
\textit{J. Appl. Phys.} \textbf{113}, 243501 (2013).

\bibitem{Gregoire2013}
J. M. Gregoire, K. Xiao, P. J. McCluskey, D. Dale, G. Cuddalorepatta, and J. J. Vlassak,
``\textit{In situ} x-ray diffraction combined with scanning AC nanocalorimetry applied to a Fe$_{0.84}$Ni$_{0.16}$ thin-film sample,''
\textit{Appl. Phys. Lett.} \textbf{102}, 201902, (2013).

\bibitem{Sun2025} P. Sun, J. Baglioni, B. Baraldi, W. Chen, D. Lideo, L. Piemontese, F. Dallari, M. Di Michiel, \& G. Monaco,
``Low-background setup for \textit{in situ} X-ray total scattering combined with fast scanning calorimetry,'' 
\textit{J. Synchrotron Radiat.} \textbf{32}, 1228-1234 (2025).

\bibitem{Melnikov2016}
A. P. Melnikov, M. Rosenthal, A. I. Rodygin, D. Doblas, D. V. Anokhin, M. Burghammer, and D. A. Ivanov,
``Re-exploring the double-melting behavior of semirigid-chain polymers with an \textit{in situ} combination of synchrotron nano-focus X-ray scattering and nanocalorimetry,''
\textit{Eur. Polym. J.} \textbf{81}, 598 (2016).

\bibitem{Rosenthal2014}
M. Rosenthal, D. Doblas, J. J. Hernandez, Y. I. Odarchenko, M. Burghammer, E. Di Cola, D. Spitzer, A. E. Antipov, L. S. Aldoshin, and D. A. Ivanov,
``High-resolution thermal imaging with a combination of nano-focus x-ray diffraction and ultra-fast chip calorimetry,''
\textit{J. Synchrotron Radiat.} \textbf{21}, 223-228 (2014).

\bibitem{Martinelli2024}
A. Martinelli, J. Baglioni, P. Sun, F. Dallari, E. Pineda, Y. Duan, 
T. Spitzbart-Silberer, F. Westermeier, M. Sprung, and G. Monaco,
``A new experimental setup for combined fast differential scanning calorimetry and X-ray photon correlation spectroscopy,''
\textit{J. Synchrotron Radiat.} \textbf{31}, 3, 557, (2024).

\bibitem{Yi2015}
F. Yi, J. B. DeLisio, M. R. Zachariah, and D. A. LaVan,
``Nanocalorimetry coupled Time-of-Flight Mass Spectrometry: Identifying evolved species during high rate thermal measurements,''
\textit{Anal. Chem.} \textbf{87}, 9740, (2015).

\bibitem{McGieson2025}
I. McGieson, S. Q. Arlington, L. R. Narayan, W. Osborn, M. D. Grapes, T. P. Weihs, M. K. Santala, D. A. LaVan, and F. Yi,
``A nanocalorimeter designed for use with high-resolution transmission electron microscopy,''
\textit{Rev. Sci. Instrum.} \textbf{96}, 103703 (2025).

\bibitem{Patel2022}
U. Patel, T. Guruswamy, A. J. Krzysko, H. Charalambous, L. Gades, K. Wiaderek, O. Quaranta, Y. Ren, A. Yakovenko, U. Ruett, and A. Miceli,
``High-resolution Compton spectroscopy using x-ray microcalorimeters,'' 
\textit{Rev. Sci. Instrum.} \textbf{93}, 113105 (2022).

\bibitem{Cole2025}
M. C. Cole, M. T. Pelly, C. V. Topping, T. Reindl, U. Waizmann, J. Weis, A. W. Rost, ``Coulomb blockade thermometry based nanocalorimetry,''
\textit{Rev. Sci. Instrum.} \textbf{96}, 073903 (2025).

\bibitem{Bachmann1972}
R. Bachmann, F. J. DiSalvo Jr., T. H. Geballe, R. L. Greene,
R. E. Howard, C. N. King, H. C. Kirsch, K. N. Lee, R. E. Schwall,
H.-U. Thomas, and R. B. Zubeck,
``Heat-capacity measurements on small samples at low temperatures,'' 
\textit{Rev. Sci. Instrum.} \textbf{43}, 205 (1972).

\bibitem{Denlinger1994}
D. W. Denlinger, E. N. Abarra, K. Allen, P. W. Rooney, M. T. Messer, S. K. Watson, and F. Hellman,
``Thin film microcalorimeter for heat capacity measurements from 1.5 to 800 K,''
\textit{Rev. Sci. Instrum.} \textbf{65}, 946-959 (1994).

\bibitem{Queen2009}
D. R. Queen and F. Hellman,
``Thin film nanocalorimeter for heat capacity measurements of 30 nm films,''
\textit{Rev. Sci. Instrum.} \textbf{80}, 063901 (2009).

\bibitem{Wickey2015}
K. J. Wickey, M. Chilcote, and E. Johnston-Halperin,
``Nanogram calorimetry using microscale suspended SiNx platforms fabricated via focused ion beam patterning,''
\textit{Rev. Sci. Instrum.} \textbf{86}, 014903 (2015).

\bibitem{Tagliati2012}
S. Tagliati, V. M. Krasnov, and A. Rydh,
``Differential membrane-based nanocalorimeter for high-resolution measurements of low-temperature specific heat,''
\textit{Rev. Sci. Instrum.} \textbf{83}, 055107 (2012).

\bibitem{Schink1981}
H. J. Schink and H.v. Lohneysen,
``Specific heat of Apiezon N grease at very low temperatures,''
\textit{Cryogenics} \textbf{21}, 591 (1981).

\bibitem{Sullivan1968}
P. F. Sullivan and G. Seidel,
``Steady-state, ac-temperature calorimetry,''
\textit{Phys. Rev.} \textbf{173}, 679 (1968).

\bibitem{Bethoux1995}
O. Béthoux, R. Brusetti, J. C. Lasjaunias, and S. Sahling,
``Au-Ge film thermometers for temperature range 30 mK-300 K,''
\textit{Cryogenics} \textbf{35}, 447 (1995).

\bibitem{SynkTeK}
Information about the SynkTek multichannel lock‑in system is available at www.synktek.com.

\bibitem{Pobell2007}
F. Pobell, ``Matter and Methods at Low Temperatures,'' (Springer, 2007), p.~291.

\bibitem{Tagliati2011}
S. Tagliati and A. Rydh,
``Absolute accuracy in membrane-based ac nanocalorimetry,''
\textit{Thermochim. Acta} \textbf{522}, 66 (2011).

\bibitem{Velichkov1992}
I.V. Velichkov,
``On the problem of thermal link resistances in a.c. calorimetry,''
\textit{Cryogenics} \textbf{32}, 285 (1992).

\bibitem{Gmelin1997}
E. Gmelin,
``Classical temperature-modulated calorimetry: A review,''
\textit{Thermochim. Acta} \textbf{304}, 1 (1997).

\bibitem{Ema1997}
K. Ema, and H. Yao,
``Some aspects of recent improvements of temperature-modulated calorimeter,''
\textit{Thermochim. Acta} \textbf{304}, 157 (1997).

\bibitem{Carbotte1990}
J. P. Carbotte,
``Properties of boson-exchange superconductors,''
\textit{Rev. Mod. Phys.} \textbf{62}, 1027 (1990).

\bibitem{Novotny1975}
V. Novotny, ``Single superconducting energy gap in pure niobium,'' 
\textit{J. Low Temp. Phys.} \textbf{18}, 147 (1975).

\bibitem{Leupold1964}
H. A. Leupold and H. A. Boorse, ``Superconducting and normal specific heats of a single crystal of niobium,''
\textit{Phys. Rev.} \textbf{134}, A1322 (1964).

\bibitem{Phillips1959}
N. E. Phillips, ``Heat Capacity of Aluminum between 0.1°K and 4.0°K,'' 
\textit{Phys. Rev.} \textbf{114}, 676 (1959).

\bibitem{Rorer1963}
D. C. Rorer, H. Meyer, and R. C. Richardson,
``Specific heat of aluminum near its superconductive transition point,"
\textit{Z. Naturforsch.} \textbf{18a}, 130--140 (1963).

\bibitem{Bouquet1998}
F. Bouquet, Étude thermodynamique de l'état mixte dans YBa$_2$Cu$_3$O$_{7-\delta}$, PhD thesis, CEA–Grenoble, France (1998).

\bibitem{Hudl2014}
M. Hudl, D. Campanini, L. Caron, V. Höglin, M. Sahlberg, P. Nordblad, and A. Rydh, “Thermodynamics around the first-order ferromagnetic phase transition of Fe$_2$P single crystals,” \textit{Phys. Rev. B} \textbf{90}, 144432 (2014).

\end{thebibliography}
\end{document}